\documentclass{article}
\usepackage{spconf,amsmath,graphicx}
\usepackage{booktabs}
\usepackage{color,soul}
\usepackage{amsfonts}
\usepackage{multicol}
\usepackage[table]{xcolor}
\usepackage[usestackEOL]{stackengine}
\usepackage[backend=bibtex,citestyle=numeric-comp,bibstyle=ieee,defernumbers=true,giveninits=true,doi=false,isbn=false,url=false,eprint=false,minbibnames=2,maxbibnames=2]{biblatex}
\AtEveryBibitem{
\clearname{editor}
}

\newcommand{\eg}{e.\,g., }
\newcommand{\ie}{i.\,e., }
\newcommand{\secvspace}{\vspace{-0.0cm}}
\newcommand{\subsecvspace}{\vspace{-0.1cm}}
\renewcommand{\tt}[1]{\selectfont #1}
\title{LA-VocE: Low-SNR Audio-visual Speech Enhancement\\ using Neural Vocoders}
\name{\parbox{\linewidth}{\centering Rodrigo Mira$^{1*}$ \qquad
Buye Xu$^{2}$ \qquad
Jacob Donley$^{2}$ \qquad
Anurag Kumar$^{2}$ \qquad
Stavros Petridis$^{1,3}$ \qquad
Vamsi Krishna Ithapu$^{2}$ \qquad
Maja Pantic$^{1,3}$\thanks{$^*$Work done during internship at Meta}}}
\address{$^{1}$iBUG, Imperial College London, UK \quad $^{2}$Meta Reality Labs Research, USA \quad $^{3}$Meta, UK}

\begin{document}
\maketitle
\ninept

\begin{abstract}
Audio-visual speech enhancement aims to extract clean speech from a noisy environment by leveraging not only the audio itself but also the target speaker's lip movements. 
This approach has been shown to yield improvements over audio-only speech enhancement, particularly for the removal of interfering speech.
Despite recent advances in speech synthesis, most audio-visual approaches continue to use spectral mapping/masking to reproduce the clean audio, often resulting in visual backbones added to existing speech enhancement architectures. 
In this work, we propose LA-VocE, a new two-stage approach that predicts mel-spectrograms from noisy audio-visual speech via a transformer-based architecture, and then converts them into waveform audio using a neural vocoder (HiFi-GAN). We train and evaluate our framework on thousands of speakers and 11+ different languages, and study our model's ability to adapt to different levels of background noise and speech interference.
Our experiments show that LA-VocE outperforms existing methods according to multiple metrics, particularly under very noisy scenarios.
\end{abstract}


\begin{figure*}[t]
\centering 
\includegraphics[width=0.9\linewidth]{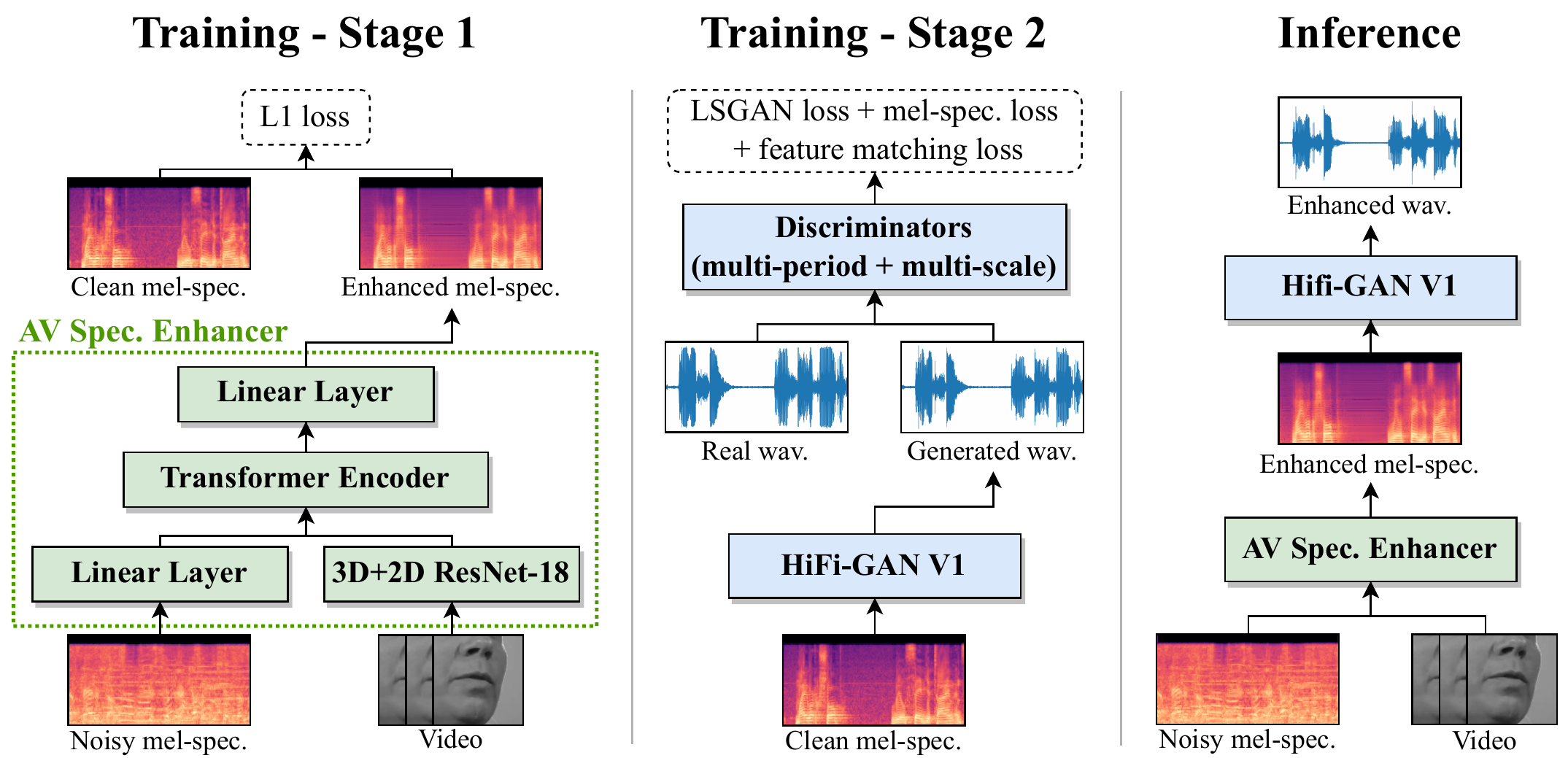}
\vspace{-0.3cm}
\caption{Summary of LA-VocE's two-stage training approach and inference procedure.}
\vspace{-0.3cm}
\label{method_fig}
\end{figure*}
\secvspace
\section{Introduction}
\label{sec:intro}
Speech enhancement, defined as the extraction of clean speech from a noisy signal, is a well-established signal processing task which has benefited greatly from the advent of deep learning~\cite{DBLP:journals/taslp/WangC18a}. Recently-proposed models excel at denoising and dereverberation~\cite{DBLP:journals/taslp/TanW20,DBLP:conf/interspeech/DefossezSA20}, but often struggle with very low signal-to-noise ratios (SNR)~\cite{DBLP:conf/interspeech/HaoSW0B19}. Furthermore, audio-only methods struggle to accurately remove background speech, as they are limited in the information they can use to distinguish it from the target signal. These limitations have drawn researchers to leverage visual cues of the target speaker's lip movements as additional supervision -- an approach known as audio-visual speech enhancement (AVSE). This can be particularly valuable for applications such as video conferencing, streaming, recording and hearing augmentation in a crowded and\,/\,or noisy environment, where the target speaker's video stream can help the model enhance their speech. This method may also be leveraged to improve speech recognition in low-SNR conditions.
Furthermore, the recent success of video-to-speech synthesis~\cite{9760273,DBLP:conf/interspeech/MiraHPSP22}, where the audio is reproduced using only silent video, highlights the importance of the visual modality and shows a promising direction for audio-visual speech enhancement in very low-SNR conditions.

Recent AVSE methods are often based on U-Nets~\cite{DBLP:conf/interspeech/GabbaySP18,DBLP:conf/icassp/PanTX021,DBLP:conf/cvpr/GaoG21}, inspired by their audio-only counterparts~\cite{DBLP:journals/taslp/LuoM19,DBLP:conf/interspeech/DefossezSA20,DBLP:journals/taslp/TanW20}, or simple convolutional networks~\cite{DBLP:journals/tetci/HouWLTCW18}, frequently combined with LSTMs~\cite{DBLP:conf/cvpr/YangMKAR22}. Existing speech enhancement models are typically combined with a video encoder which extracts visual features and concatenates them with the acoustic features to perform audio-visual enhancement. These approaches draw from speech enhancement literature, but fail to leverage state-of-the-art audio-visual encoders~\cite{DBLP:conf/icassp/0001PP21a,DBLP:conf/iclr/ShiHLM22}. Most methods estimate (either directly or via a mask) the magnitude and phase of the clean spectrogram, which are converted into waveform using the inverse Short-Time Fourier Transform (iSTFT)~\cite{DBLP:conf/interspeech/GabbaySP18,DBLP:journals/tetci/HouWLTCW18,DBLP:conf/cvpr/GaoG21}, while others attempt to perform enhancement in the time domain directly~\cite{DBLP:conf/icassp/PanTX021}. Both of these reconstruction techniques rely on very accurate predictions, which can be difficult to achieve, especially in low-SNR environments where audio supervision is unreliable. Recent works in audio-only \cite{DBLP:journals/taslp/DuZH20,DBLP:conf/interspeech/LiY20} and audio-visual ~\cite{DBLP:conf/cvpr/YangMKAR22} speech enhancement have introduced neural vocoders as an alternative synthesis method, but choose to focus on high-SNR scenarios where this reconstruction technique is likely to have a lesser impact. Alternatively, new works introduce neural codecs~\cite{DBLP:journals/taslp/ZeghidourLOST22} for waveform synthesis but focus heavily on achieving compressed representations, which is not a priority for most speech enhancement frameworks.

To address these shortcomings, we propose a new two-stage approach for audio-visual speech enhancement entitled \textbf{L}ow-SNR \textbf{A}udio-visual \textbf{Voc}oder-based Speech \textbf{E}nhancement  
(LA-VocE). First, we train an audio-visual spectrogram enhancer, which receives noisy speech and video of the cropped mouth, and aims to predict a clean spectrogram. This model features a ResNet-18-based visual encoder~\cite{DBLP:conf/cvpr/HeZRS16} and a large transformer encoder~\cite{DBLP:conf/nips/VaswaniSPUJGKP17} to model the temporal correlations in the audio-visual features, and is trained using an L1 loss between the real and predicted mel-spectrogram magnitudes. We then train a neural vocoder (HiFi-GAN V1~\cite{NEURIPS2020_c5d73680}) to predict waveform audio from clean mel-spectrograms on the same corpus. This fully convolutional model is trained using a mixture of adversarial and comparative losses, with an ensemble of eight discriminators operating on multiple periods and scales. During inference, the enhancer and the vocoder are combined to perform end-to-end audio-visual speech enhancement.

Our contributions are as follows: \textbf{(1)} We present a new audio-visual speech enhancement approach that combines a transformer-based spectrogram enhancer with our version of HiFi-GAN V1. \textbf{(2)} We train our model to remove background noise and speech on the challenging AVSpeech dataset. 
\textbf{(3)} We compare our approach with previous state-of-the-art models, and show that it significantly outperforms all methods across all metrics and noise conditions.
\textbf{(4)} We study our model's ability to generate clean audio for varying levels of noise and interference and find that it consistently achieves improvements in speech intelligibility. 
\textbf{(5)} We measure our trained vocoder's effectiveness against other spectrogram inversion approaches and observe that it significantly outperforms other methods.
\secvspace
\section{Methodology}
\secvspace
\subsection{Audio-visual spectral enhancement}
\subsecvspace
LA-VocE is summarized in Fig.~\ref{method_fig}. The first stage in our framework consists of training an audio-visual spectrogram enhancer. This model extracts visual features using a 2D ResNet-18~\cite{DBLP:conf/cvpr/HeZRS16} with a 3D convolutional stem (as in~\cite{DBLP:conf/icassp/PetridisSMCTP18,9760273,DBLP:journals/corr/abs-2202-13084,DBLP:conf/interspeech/MiraHPSP22}), and acoustic features using a single linear layer.
The video features are then upsampled (via nearest neighbors interpolation) to match the audio features' frame rate, and the features from the two modalities are concatenated along the channel dimension. The fused audio-visual features are fed into the transformer encoder~\cite{DBLP:conf/nips/VaswaniSPUJGKP17} - the largest component in the network. This module comprises an initial embedding layer, with a linear layer followed by relative positional encoding~\cite{DBLP:conf/acl/DaiYYCLS19}, and 12 transformer encoder blocks, where the attention dimension, feedforward dimension, and the number of attention heads are 768, 3072, and 12, respectively. Finally, these features are decoded via a linear layer into the predicted mel-frequency spectrogram. We train the model by applying an L1 Loss:
 \begin{equation}
     \mathcal{L}_{1} = \| s_{clean} - E(s_{noisy},v) \|_1,
 \end{equation}
 where $s_{clean}$ and $s_{noisy}$ are the clean and noisy mel-spectrograms, respectively, $v$ is the video of the speaker's lip movements and $E$ is our audio-visual spectrogram enhancer.
\subsecvspace
\subsection{Waveform synthesis}
\subsecvspace
The second stage in our method involves training a neural vocoder to convert the enhanced spectrograms into waveform audio. We use HiFi-GAN~\cite{NEURIPS2020_c5d73680}, which upsamples the spectrogram gradually using a set of transposed convolutions. In particular, we opt for HiFi-GAN V1, the largest HiFi-GAN variant, which features 12 ResBlocks with hidden size 512, amounting to 13.92 million parameters. As proposed in~\cite{NEURIPS2020_c5d73680}, HiFi-GAN is trained via a multi-period discriminator (MPD), composed of five convolutional sub-discriminators which analyze the waveform along different periods (\ie every 2, 3, 5, 7 and 11 samples), and a multi-scale discriminator (MSD), consisting of one sub-discriminator for the raw audio and two sub-discriminators that receive downsampled versions of the same waveform (via 2$\times$ and 4$\times$ average pooling). Our training objective (as in the original HiFi-GAN) combines the Least Squares Generative Adversarial Network (LSGAN) loss~\cite{DBLP:conf/iccv/MaoLXLWS17} with an L1 loss on the mel-spectrogram magnitudes and a feature matching loss~\cite{DBLP:conf/icml/LarsenSLW16}: 
\begin{equation}
    \mathcal{L}_{G} = \alpha_{1} \mathcal{L}_{G_{adv}} + \alpha_{2} \mathcal{L}_{spec} + \alpha_{3} \mathcal{L}_{FM},
\end{equation}
\begin{equation}
    \mathcal{L}_{G_{adv}} = \sum_{i=1}^{N_D} (D_i(G(s_{clean})) - 1)^2,
\end{equation}
\begin{equation}
    \mathcal{L}_{spec} = \| m(w_{clean})- m(G(s_{clean})) \|_1,
\end{equation}
\begin{equation}
    \mathcal{L}_{FM} = \sum_{i=1}^{N_D} \sum_{l=1}^{N_{L_{i}}} \frac{\| D_{i}^{l}(w_{clean}) - D_{i}^{l}(G(s_{clean})) \|_1} {d_{i}^{l}},
\end{equation}
\begin{equation}
    \mathcal{L}_{D} = \sum_{i=1}^{N_D} (D_i(w_{clean}) - 1)^2 + D_i(G(s_{clean}))^2,
\end{equation}
where $\mathcal{L}_G$ is the generator loss, $\mathcal{L}_D$ is the discriminator loss, $\mathcal{L}_{G_{adv}}$ is the generator's adversarial loss, $\mathcal{L}_{spec}$ is the mel-spectrogram loss, $\mathcal{L}_{FM}$ is the feature matching loss, $G$ is the generator (HiFi-GAN V1), $D_i$ is the $i$-th discriminator, $N_D$ is the number of discriminators, $w_{clean}$ is the clean waveform, $m$ is the function that computes the mel-spectrogram, $N_{L_{i}}$ is the number of layers in discriminator $i$, and $D_{i}^{l}$ and $d_{i}^{l}$ refer to the features extracted from layer $l$\,/\,discriminator $i$ and their dimension, respectively. Loss coefficients $\alpha_1$, $\alpha_2$ and $\alpha_3$ are set to 1, 45, and 2, respectively, as in \cite{NEURIPS2020_c5d73680}.
After both stages of training, the spectrogram enhancer and neural vocoder are combined during inference, as shown in Fig.~\ref{method_fig}. 
\secvspace
\section{Experimental Setup}
\secvspace
\subsection{Datasets, pre-processing, and augmentation}
Our experiments focus on AVSpeech~\cite{DBLP:journals/tog/EphratMLDWHFR18}, one of the largest publicly available audio-visual speech datasets. It contains $\sim$4,700 hours of video, featuring $\sim$150,000 different subjects and 11+ languages. 
The scale and heterogeneity of the data make for a substantially more challenging than many commonly-used corpora such as GRID~\cite{grid,DBLP:conf/interspeech/GabbaySP18,DBLP:conf/cvpr/YangMKAR22} or Facestar~\cite{DBLP:conf/cvpr/YangMKAR22}, which are recorded in studios.

We sample background noise from the Deep Noise Supression challenge~\cite{DBLP:conf/interspeech/ReddyGCBCDMAABR20} noise dataset. It contains roughly 70,000 noise clips, amounting to around 150 classes, ranging from music to machine sounds.
Both datasets are split into training, validation and testing sets using a 80 -- 10 -- 10\,\% ratio. Due to the computational cost of computing the evaluation metrics, we randomly sample 1\,\% of the AVSpeech testing set, amounting to 1552 samples, and use this as the evaluation set for our experiments.
We add two types of corruption to the clean speech: background noise (denoted `noise') and background speech (denoted `interference'). The corruption level is controlled by the Signal-to-Noise Ratio (SNR) and the Signal-to-Interference Ratio (SIR):
\vspace{-0.1cm}
\begin{align}
\text{SNR} =  \frac{P_{signal}}{P_{noise}},  & \qquad \qquad \quad \text{SIR} = \frac{P_{signal}}{P_{interference}},
\end{align}
\noindent where P refers to the power of each waveform. The interfering speech is also obtained from AVSpeech. During training, the SNR and SIR are independently randomly sampled between 5 and -15 dB. 
The number of background noises and interfering speakers in each sample varies randomly from 1 to 5 and 1 to 3, respectively. During validation, we propose three different noise conditions to compare with other methods, ranging from least to most noisy. Noise conditions 1 (low), 2 (medium), and 3 (high) feature 1, 3, and 5 background noises at 0, -5, and -10 dB SNR, and 1, 2, and 3 background speakers at 0, -5, and -10 dB SIR, respectively.

The noisy and clean signals are normalized via peak normalization, and are converted into log-scale mel-spectrograms using the following parameters: frequency bin size and Hann window size 1024, hop size 256 and 80 mel bands. The audio sampling rate is 16 kHz and the video frame rate is 25 frames per second (fps). To model the speaker's lip movements, we extract the $96 \times 96$ grayscale mouth Region Of Interest (ROI) from each video, following~\cite{DBLP:journals/corr/abs-2202-13084,DBLP:conf/interspeech/MiraHPSP22}.
To augment our data, we apply random cropping, random horizontal flipping, random erasing, and time-masking, as in~\cite{DBLP:conf/interspeech/MiraHPSP22}.
\subsecvspace
\subsection{Evaluation metrics}
\subsecvspace
We evaluate our results using a set of five objective speech metrics. To measure speech quality, we use Mean Cepstral Distance (MCD)~\cite{407206}, the wideband version of Perceptual Evaluation of
Speech Quality 
(PESQ-WB)~\cite{DBLP:conf/icassp/RixBHH01}, and Virtual Speech Quality Objective Listener (ViSQOL)~\cite{DBLP:conf/qomex/ChinenLSGOH20}. To measure intelligibility we use Short-Time Objective Intelligibility (STOI)~\cite{DBLP:conf/icassp/TaalHHJ10} and its extended version ESTOI~\cite{DBLP:journals/taslp/JensenT16}. Finally, in our spectrogram inversion comparison, we  also measure the mean squared error between the STFT magnitudes of each signal and refer to this as Spec. MSE.
Following other works~\cite{DBLP:journals/taslp/LuoM19,DBLP:conf/icassp/PanTX021}, we denote improvements between noisy and enhanced speech metrics with the lowercase `i', \eg PESQ-WB\,i.
\subsecvspace
\subsection{Comparison models}
\subsecvspace
We compare our results with two recent AVSE models: VisualVoice~\cite{DBLP:conf/cvpr/GaoG21}, a complex spectral masking approach originally proposed for speech separation that we adapt to perform enhancement, and Multi-modal Speaker Extraction (MuSE)~\cite{DBLP:conf/icassp/PanTX021}, a feature masking approach based on Conv-TasNet~\cite{DBLP:journals/taslp/LuoM19}. To provide a broader comparison with other reconstruction techniques, we also adapt two recent speech enhancement models for AVSE - Gated Convolutional Recurrent Network (GCRN)~\cite{DBLP:journals/taslp/TanW20} and Demucs~\cite{DBLP:conf/interspeech/DefossezSA20}. We achieve this by adding a visual stream (3D front-end + ResNet-18, as in our model) which encodes the video into temporal features that are concatenated with the audio features from the original audio encoder (preceding the LSTM/GLSTM). We refer to these models as AV-GCRN and AV-Demucs. We also compare with the original audio-only GCRN and an audio-only version of LA-VocE to highlight the importance of the visual stream. All models are implemented based on official open-source code. 
\subsecvspace
\subsection{Training details}
\subsecvspace
We train our spectrogram enhancer for 150 epochs using AdamW~\cite{DBLP:conf/iclr/LoshchilovH19} with learning rate $7\times10^{-4}$, $\beta_1 = 0.9$, $\beta_2 = 0.98$ and weight decay $3\times10^{-2}$. We increase the learning rate for the first 15 epochs using linear warmup, and then apply a cosine decay schedule~\cite{DBLP:conf/iclr/LoshchilovH17}. 
To train MuSE, we replace the original SI-SDR objective~\cite{DBLP:conf/icassp/PanTX021} with the loss from Demucs (L1 + multi-resolution STFT~\cite{DBLP:conf/interspeech/DefossezSA20}), as we find this increases training stability and  yields better results. We train an audio-only version of LA-VocE by removing the visual encoder and changing the attention dimension and the number of heads in the transformer to 256 and 8, respectively.  We train HiFi-GAN for roughly 1 million iterations on AVSpeech using AdamW with learning rate $2\times10^{-4}$, $\beta_1 = 0.8$, $\beta_2 = 0.99$ and weight decay $1\times10^{-2}$, decaying the learning rate by a factor of 0.999 every epoch. 
\secvspace
\section{Results}
\secvspace
\subsection{Comparison with other works}
\subsecvspace
We compare with previous state-of-the-art methods in Table~\ref{comparison_table}, and present a demo of these results on our project website\footnote{\url{https://sites.google.com/view/la-voce-avse}}. For noise condition 1, LA-VocE outperforms other approaches in quality and intelligibility, achieving significant improvements across all metrics. Indeed, even in this less noisy scenario, our vocoder-based approach is able to reproduce the target speech more accurately than mask-based methods such as MuSE~\cite{DBLP:conf/icassp/PanTX021} and VisualVoice~\cite{DBLP:conf/cvpr/GaoG21}, which are designed for separation with one to two background speakers. Previous AVSE methods yield decreased improvements for noise condition 2, particularly for speech quality metrics such as PESQ and ViSQOL, while LA-VocE yields significant gains in quality and especially intelligibility, as indicated by ESTOI\,i. This shows that despite identical training conditions, previous methods adapt poorly to lower SNR\,/\,SIR conditions compared to our new model. 

Finally, on the noisiest scenario (noise condition 3), it is clear that other audio-visual methods, including mapping-based approaches (AV-GCRN~\cite{DBLP:journals/taslp/TanW20} and AV-Demucs~\cite{DBLP:conf/interspeech/DefossezSA20}), are unable to increase speech quality, achieving effectively no improvement on PESQ-WB and small increases on other metrics. LA-VocE, on the other hand, can still achieve significant gains in all metrics, indicating that it is substantially more robust to extremely low-SNR scenarios. Notably, both audio-only models (GCRN~\cite{DBLP:journals/taslp/TanW20} and LA-VocE) yield poor results in all scenarios - without visual information, these models cannot accurately distinguish target speech from background speech.

\subsecvspace
\subsection{Noise and interference study}
\subsecvspace
\begin{table}[t]
  \centering
  \caption{Comparison between LA-VocE and other speech enhancement methods for different noise conditions. In the second column, ``A'' and ``AV'' stand for audio-only and audio-visual, respectively.}
  \vspace{0.1cm}
  \resizebox{\linewidth}{!}{
  \begin{tabular}{lc|ccccc}
    \toprule
    \textbf{Method} & \textbf{Input} & \textbf{MCD\,i} $\downarrow$ & \textbf{PESQ-WB\,i} $\uparrow$ &  \textbf{ViSQOL\,i} $\uparrow$ & \textbf{STOI\,i} $\uparrow$ & \textbf{ESTOI\,i} $\uparrow$ \\
    \midrule
    \midrule
    \rowcolor{lightgray!30}
    \multicolumn{7}{l}{\textbf{Noise condition 1 (1 background noise at 0 dB SNR + 1 interfering speaker at 0 dB SIR)}} \vspace{.5em}\\
    GCRN~\cite{DBLP:journals/taslp/TanW20} & A &\tt{ 0.410} &  \tt{ 0.044}  & \tt{ 0.093}& \tt{-0.052} & \tt{-0.038}\\
    AV-GCRN~\cite{DBLP:journals/taslp/TanW20} & AV &\tt{-1.193} &  \tt{ 0.394}  & \tt{ 0.499}& \tt{ 0.220} & \tt{ 0.235}\\
    AV-Demucs~\cite{DBLP:conf/interspeech/DefossezSA20} & AV & \tt{-5.581} & \tt{ 0.738} & \tt{ 0.688}& \tt{ 0.270} & \tt{ 0.298}\\
    MuSE~\cite{DBLP:conf/icassp/PanTX021} & AV & \tt{-5.528} & \tt{ 0.787} & \tt{ 0.679}& \tt{ 0.276} & \tt{ 0.299} \\
    VisualVoice~\cite{DBLP:conf/cvpr/GaoG21} & AV & \tt{-3.781} & \tt{ 0.606}  & \tt{ 0.645}& \tt{ 0.249} & \tt{ 0.270}\\
    LA-VocE (audio-only) & A & \tt{-3.189} & \tt{0.248} & \tt{0.135} & \tt{0.055} & \tt{0.047} \\
    LA-VocE & AV & \textbf{\tt{-6.653}} & \textbf{\tt{ 0.931}} & \textbf{\tt{ 1.100}}& \textbf{\tt{ 0.294}} & \textbf{\tt{ 0.333}} \\
    \midrule
    \rowcolor{lightgray!30}
    \multicolumn{7}{l}{\textbf{Noise condition 2 (3 background noises at -5 dB SNR + 2 interfering speakers at -5 dB SIR)}} \vspace{.5em}\\
    GCRN~\cite{DBLP:journals/taslp/TanW20} & A &\tt{-0.416} &  \tt{-0.010} & \tt{ 0.163}& \tt{-0.015} & \tt{-0.015} \\
    AV-GCRN~\cite{DBLP:journals/taslp/TanW20} & AV &\tt{-1.354} &  \tt{ 0.096} & \tt{ 0.398}& \tt{ 0.234} & \tt{ 0.214} \\
    AV-Demucs~\cite{DBLP:conf/interspeech/DefossezSA20} & AV & \tt{-5.548} & \tt{ 0.274} &\tt{ 0.426}& \tt{ 0.308} & \tt{ 0.300} \\
    MuSE~\cite{DBLP:conf/icassp/PanTX021} & AV & \tt{-5.314} & \tt{ 0.297} & \tt{ 0.409}& \tt{ 0.308} & \tt{ 0.289} \\
    VisualVoice~\cite{DBLP:conf/cvpr/GaoG21} & AV  & \tt{-3.388} & \tt{ 0.164} & \tt{ 0.367}& \tt{ 0.253} & \tt{ 0.237} \\
    LA-VocE (audio-only) & A & \tt{-2.817} & \tt{0.056} & \tt{0.087} & \tt{0.066} & \tt{0.043} \\
    LA-VocE & AV & \textbf{\tt{-6.863}} & \textbf{\tt{ 0.511}} & \textbf{\tt{ 0.700}}& \textbf{\tt{ 0.379}} & \textbf{\tt{ 0.397}} \\
    \midrule
    \rowcolor{lightgray!30}
    \multicolumn{7}{l}{\textbf{Noise condition 3 (5 background noises at -10 dB SNR + 3 interfering speakers at -10 dB SIR)}} \vspace{.5em}\\
    GCRN~\cite{DBLP:journals/taslp/TanW20} & A & \tt{-0.414} & \tt{-0.015} & \tt{ 0.210}& \tt{-0.020} & \tt{-0.005} \\
    AV-GCRN~\cite{DBLP:journals/taslp/TanW20} & AV & \tt{-1.263} & \tt{-0.043} & \tt{ 0.217}& \tt{ 0.171} & \tt{ 0.139} \\
    AV-Demucs~\cite{DBLP:conf/interspeech/DefossezSA20} & AV & \tt{-4.866} & \tt{ 0.013}  & \tt{ 0.298}& \tt{ 0.262} & \tt{ 0.230}\\
    MuSE~\cite{DBLP:conf/icassp/PanTX021} & AV & \tt{-4.185} & \tt{ 0.011} & \tt{ 0.242}& \tt{ 0.231} & \tt{ 0.182} \\
    VisualVoice~\cite{DBLP:conf/cvpr/GaoG21} & AV &\tt{-2.518} & \tt{-0.045} & \tt{ 0.248}& \tt{ 0.181} & \tt{ 0.160} \\
    LA-VocE (audio-only) & A & \tt{-1.982} & \tt{-0.015} & \tt{0.073} & \tt{0.032} & \tt{0.008} \\
    LA-VocE & AV & \textbf{\tt{-6.170}} & \textbf{\tt{ 0.159}} & \textbf{\tt{ 0.447}}& \textbf{\tt{ 0.371}} & \textbf{\tt{ 0.358}} \\
    \bottomrule
  \end{tabular}
  }
  \vspace{-0.3cm}
  \label{comparison_table}
\end{table}
\begin{table}[t]
  \centering
  \caption{LA-VocE's performance for different SNR\,/\,SIR conditions with 3 background noises and 2 interfering speakers.}
  \vspace{0.1cm}
  \resizebox{\linewidth}{!}{
  \begin{tabular}{cc|ccccc|ccccc}
    \toprule
    \multicolumn{2}{c}{} & \multicolumn{5}{c}{\textbf{PESQ-WB\,i $\uparrow$}} & \multicolumn{5}{c}{\textbf{ESTOI\,i $\uparrow$}} \\
    \midrule
    \midrule
    \multicolumn{2}{c}{\textbf{SNR (dB)}} & \textbf{5} & \textbf{0} & \textbf{-5} & \textbf{-10} & \textbf{-15} & \textbf{5} & \textbf{0} & \textbf{-5} & \textbf{-10} & \textbf{-15}\\
    \midrule
    &\textbf{5} & \cellcolor{lightgray!100.0}\tt{0.970} & \cellcolor{lightgray!90.30927835051546}\tt{0.876} & \cellcolor{lightgray!73.71134020618557}\tt{0.715} & \cellcolor{lightgray!50.103092783505154}\tt{0.486} & \cellcolor{lightgray!25.257731958762886}\tt{0.245} & \cellcolor{lightgray!60.72234762979684}\tt{0.269} & \cellcolor{lightgray!71.33182844243792}\tt{0.316} & \cellcolor{lightgray!80.36117381489842}\tt{0.356} & \cellcolor{lightgray!84.65011286681715}\tt{0.375} & \cellcolor{lightgray!81.71557562076748}\tt{0.362} \\
&\textbf{0} & \cellcolor{lightgray!93.1958762886598}\tt{0.904} & \cellcolor{lightgray!81.95876288659794}\tt{0.795} & \cellcolor{lightgray!64.94845360824742}\tt{0.630} & \cellcolor{lightgray!42.371134020618555}\tt{0.411} & \cellcolor{lightgray!21.649484536082475}\tt{0.210} & \cellcolor{lightgray!73.81489841986456}\tt{0.327} & \cellcolor{lightgray!79.90970654627539}\tt{0.354} & \cellcolor{lightgray!84.65011286681715}\tt{0.375} & \cellcolor{lightgray!85.32731376975168}\tt{0.378} & \cellcolor{lightgray!80.13544018058691}\tt{0.355} \\
\smash{\rotatebox[origin=c]{90}{\textbf{SIR (dB)}}}&\textbf{-5} & \cellcolor{lightgray!81.34020618556701}\tt{0.789} & \cellcolor{lightgray!70.00000000000001}\tt{0.679} & \cellcolor{lightgray!52.577319587628864}\tt{0.511} & \cellcolor{lightgray!32.886597938144334}\tt{0.319} & \cellcolor{lightgray!14.020618556701033}\tt{0.136} & \cellcolor{lightgray!87.13318284424379}\tt{0.386} & \cellcolor{lightgray!88.93905191873588}\tt{0.394} & \cellcolor{lightgray!89.61625282167043}\tt{0.397} & \cellcolor{lightgray!86.45598194130925}\tt{0.383} & \cellcolor{lightgray!78.78103837471782}\tt{0.349} \\
&\textbf{-10} & \cellcolor{lightgray!63.608247422680414}\tt{0.617} & \cellcolor{lightgray!53.917525773195884}\tt{0.523} & \cellcolor{lightgray!41.75257731958763}\tt{0.405} & \cellcolor{lightgray!25.567010309278352}\tt{0.248} & \cellcolor{lightgray!9.484536082474227}\tt{0.092} & \cellcolor{lightgray!96.83972911963882}\tt{0.429} & \cellcolor{lightgray!96.16252821670429}\tt{0.426} & \cellcolor{lightgray!93.45372460496614}\tt{0.414} & \cellcolor{lightgray!87.58465011286683}\tt{0.388} & \cellcolor{lightgray!77.65237020316027}\tt{0.344} \\
&\textbf{-15} & \cellcolor{lightgray!45.15463917525773}\tt{0.438} & \cellcolor{lightgray!39.48453608247422}\tt{0.383} & \cellcolor{lightgray!29.79381443298969}\tt{0.289} & \cellcolor{lightgray!20.103092783505154}\tt{0.195} & \cellcolor{lightgray!8.350515463917526}\tt{0.081} & \cellcolor{lightgray!99.99999999999999}\tt{0.443} & \cellcolor{lightgray!97.74266365688487}\tt{0.433} & \cellcolor{lightgray!93.45372460496614}\tt{0.414} & \cellcolor{lightgray!86.00451467268623}\tt{0.381} & \cellcolor{lightgray!74.49209932279909}\tt{0.330} \\
    
    \bottomrule
  \end{tabular}
  }
  \label{snr_table}
  \vspace{-0.3cm}
\end{table}
\begin{table}[t]
  \centering
  \caption{LA-VocE's performance for different numbers of background noises and interfering speakers (-5 dB SNR\,/\,SIR).}
  \vspace{0.1cm}
  \resizebox{\linewidth}{!}{
  \begin{tabular}{cc|ccccc|ccccc}
    \toprule
    \multicolumn{2}{c}{} & \multicolumn{5}{c}{\textbf{PESQ-WB\,i $\uparrow$}} & \multicolumn{5}{c}{\textbf{ESTOI\,i $\uparrow$}} \\
    \midrule
    \midrule
    \multicolumn{2}{c}{\textbf{\textbf{\# noises}}}      & \textbf{1} & \textbf{2} & \textbf{3} & \textbf{4} & \textbf{5} & \textbf{1} & \textbf{2} & \textbf{3} & \textbf{4} & \textbf{5}\\
    \midrule
    &\textbf{1} & \cellcolor{lightgray!100.0}\tt{0.709} & \cellcolor{lightgray!90.55007052186178}\tt{0.642} & \cellcolor{lightgray!84.7672778561354}\tt{0.601} & \cellcolor{lightgray!81.80535966149506}\tt{0.580} & \cellcolor{lightgray!78.56135401974613}\tt{0.557} & \cellcolor{lightgray!98.01980198019801}\tt{0.396} & \cellcolor{lightgray!99.5049504950495}\tt{0.402} & \cellcolor{lightgray!100.00000000000001}\tt{0.404} & \cellcolor{lightgray!100.00000000000001}\tt{0.404} & \cellcolor{lightgray!99.75247524752476}\tt{0.403} \\
\smash{\rotatebox[origin=c]{90}{\textbf{\# spk.}}}&\textbf{2} & \cellcolor{lightgray!84.9083215796897}\tt{0.602} & \cellcolor{lightgray!77.99717912552893}\tt{0.553} & \cellcolor{lightgray!71.93229901269395}\tt{0.511} & \cellcolor{lightgray!70.09873060648802}\tt{0.497} & \cellcolor{lightgray!67.98307475317348}\tt{0.482} & \cellcolor{lightgray!98.01980198019801}\tt{0.396} & \cellcolor{lightgray!98.51485148514853}\tt{0.398} & \cellcolor{lightgray!98.26732673267327}\tt{0.397} & \cellcolor{lightgray!97.77227722772277}\tt{0.395} & \cellcolor{lightgray!97.27722772277228}\tt{0.393} \\
&\textbf{3} & \cellcolor{lightgray!76.0225669957687}\tt{0.539} & \cellcolor{lightgray!69.1114245416079}\tt{0.490} & \cellcolor{lightgray!65.16220028208745}\tt{0.462} & \cellcolor{lightgray!64.17489421720734}\tt{0.455} & \cellcolor{lightgray!60.78984485190409}\tt{0.431} & \cellcolor{lightgray!96.53465346534652}\tt{0.390} & \cellcolor{lightgray!96.53465346534652}\tt{0.390} & \cellcolor{lightgray!96.03960396039605}\tt{0.388} & \cellcolor{lightgray!95.79207920792079}\tt{0.387} & \cellcolor{lightgray!95.04950495049503}\tt{0.384} \\
    
    \bottomrule
  \end{tabular}
  }
  \label{nspk_table}
  \vspace{-0.3cm}
\end{table}
\begin{table}[t]
  \centering
  \caption{Comparison between different spectrogram inversion methods for LA-VocE (noise condition 2). In the upper row, ``Train. corp.'' stands for training corpus.}
  \vspace{0.1cm}
  \resizebox{\linewidth}{!}{
  \begin{tabular}{@{\hskip3pt}l@{\hskip3pt}c@{\hskip3pt}|@{\hskip3pt}c@{\hskip6pt}c@{\hskip6pt}c@{\hskip6pt}c@{\hskip6pt}c@{\hskip3pt}c@{\hskip3pt}c@{\hskip3pt}}
    \toprule
    \textbf{Method} & \textbf{Train. corp.} & \textbf{MCD\,i} $\downarrow$ & \textbf{PESQ-WB\,i} $\uparrow$ & \textbf{ViSQOL\,i} $\uparrow$ & \textbf{STOI\,i} $\uparrow$ & \textbf{ESTOI\,i} $\uparrow$ & \textbf{Spec. MSE\,i} $\downarrow$\\ 
    \midrule
    \midrule
    Griffin-Lim~\cite{1164317} & - & -6.805 & 0.333 & \textbf{0.806}& 0.311 & 0.318 & -7.855 \\ 
    Noisy phase & - & \tt{-6.640} & \tt{0.461} & \tt{0.721} & \tt{0.305} & \tt{0.310} & -7.901 \\ 
    HiFi-GAN~\cite{NEURIPS2020_c5d73680} & VCTK & -6.570 & 0.384& 0.655 & 0.374 & 0.388 & -7.773 \\ 
    HiFi-GAN~\cite{NEURIPS2020_c5d73680} & LJSpeech & -6.601 & 0.432 & 0.670& 0.370 & 0.382 & -7.825 \\ 
    HiFi-GAN~\cite{NEURIPS2020_c5d73680} & AVSpeech & \textbf{-6.863} & \textbf{0.511} & 0.700 & \textbf{0.379} & \textbf{0.397} & \textbf{-7.939} \\ 
    
    \bottomrule
  \end{tabular}
  }
  \label{vocoder_table}
  \vspace{-0.3cm}
\end{table}
We study our model's performance in Table~\ref{snr_table} by varying the SNR and SIR between 5 dB and -15 dB (as in training), while keeping the number of background noises and interfering speakers fixed at 3 and 2, respectively. On the left, we can see that PESQ-WB\,i peaks for higher SNR\,/\,SIR conditions and deteriorates as the noise and interference increase. This suggests that the model excels at improving speech quality for higher SNR\,/\,SIR, even with the higher PESQ baseline, but struggles to achieve substantial gains when the environment becomes too noisy. On the other hand, ESTOI\,i is substantially more consistent across all conditions, and is in fact higher for -15 dB SNR\,/\,SIR than it is for 5 dB SNR\,/\,SIR. Indeed, LA-VocE achieves impressive improvements in intelligibility even for -15 dB SNR\,/\,SIR, where the target speech is entirely imperceptible for human listeners. This is consistent with our perceptual evaluation - LA-VocE consistently produces intelligible audio despite the noticeable artifacts for lower SNR\,/\,SIRs. 

Remarkably, LA-VocE performs better for lower SIRs compared to lower SNRs, \eg 5 dB SNR\,/\,-15 dB SIR substantially outperforms -15 dB SNR\,/\,5 dB SIR on both metrics. This disparity is likely due to the nature of these two signals. Speech typically has a consistent frequency range, and often contains gaps that the model can easily exploit, while noise is substantially more heterogeneous, ranging from impulses to continuous noises, presenting a greater denoising challenge. 
We also evaluate our model's ability to perform enhancement under multiple noise sources and background speakers in Table~\ref{nspk_table}, keeping the SNR\,/\,SIR at -5 dB. Unsurprisingly, we find that the best PESQ-WB\,i is achieved with 1 noise and 1 speaker, and becomes worse as they are increased. 
While it is expected that increasing the number of sources will increase the complexity of the background noise, therefore making the enhancement task more difficult, we hypothesize that the sharper drop in performance when increasing the number of speakers is related to the temporal and spectral gaps in the interference. A single stream of speech will contain pauses that will ease denoising, but these disappear as we increase the number of speakers, resembling continuous noise.
\subsecvspace
\subsection{Spectrogram inversion comparison}
\subsecvspace
Finally, we compare our trained HiFi-GAN with other spectrogram inversion methods in Table~\ref{vocoder_table}. We observe that our HiFi-GAN achieves better performance than existing pre-trained models\footnote{ \url{https://github.com/jik876/hifi-gan}} (trained on VCTK~\cite{VCTK} and LJSpeech~\cite{ljspeech17}, as presented in \cite{NEURIPS2020_c5d73680}) on all six metrics, highlighting the importance of training our own vocoder on AVSpeech, rather than applying a publicly available pre-trained model as in \cite{DBLP:conf/interspeech/MiraHPSP22}. We also compare with Griffin-Lim~\cite{1164317}, a commonly-used spectrogram inversion algorithm, and experiment by applying iSTFT using the phase from the noisy input to reconstruct the waveform, as proposed in \cite{DBLP:conf/interspeech/GabbaySP18,DBLP:journals/tetci/HouWLTCW18}. In our experiments, both methods consistently produce artifacts that make the resulting waveforms sound noticeably more robotic than those produced by neural vocoders (this is particularly noticeable for Griffin-Lim). We show that these inversion methods yield significant drops in PESQ-WB\,i, STOI\,i, and ESTOI\,i, but surprisingly achieve competitive MCD\,i and Spec. MSE\,i performance, and substantially better ViSQOL\,i. This inconsistency likely implies that these three metrics are less sensitive to the specific artifacts introduced by these phase estimation strategies, and emphasizes the need for multiple evaluation metrics when evaluating synthesized speech.
\secvspace
\section{Conclusion}
\secvspace
In this paper, we propose LA-VocE, a new framework for audio-visual speech enhancement under low-SNR conditions. Our method consists of two stages of training: audio-visual spectral enhancement via a transformer-based encoder, and waveform synthesis via HiFi-GAN. We train our model on thousands of hours of multilingual audio-visual speech, and find that it significantly outperforms previous state-of-the-art AVSE approaches, particularly for higher noise conditions. We study LA-VocE's performance under varying levels of noise and interference, showing that even in the noisiest scenarios our vocoder-based approach can achieve large improvements in speech intelligibility. Finally, we compare our vocoder with existing spectrogram inversion methods, highlighting the importance of training our own HiFi-GAN.
In the future, we believe it would be promising to adapt our architecture for real-time synthesis, which would enable speech enhancement in live video streams.
\secvspace
\section{Acknowledgements}
\secvspace
Only non-Meta authors conducted any of the dataset pre-processing (no dataset pre-processing took place on Meta’s servers or facilities).
\clearpage
\centering\section*{References}
\vspace{-0.2cm}
\setlength\bibitemsep{0.3\itemsep}
\AtNextBibliography{\ninept}
\printbibliography[heading=none]

\end{document}